\documentclass[letter,preprint]{jpsj2}
\usepackage[]{graphicx,color}

\def\runauthor{Yoshihiro Shimomura}

\title{%
Spin Dodecamer Formation \\
in the Double-Exchange Spin Ice Model
}

\author{%
Yoshihiro {\sc Shimomura}\thanks{E-mail address: shimomura@phys.aoyama.ac.jp},
Shin {\sc Miyahara} and Nobuo {\sc Furukawa} 
}

\inst{%
Department of Physics and Mathematics, Aoyama Gakuin University, Sagamihara 229-8558, Japan
}

\recdate{\today}

\abst{%
We investigated the double-exchange spin ice (DESI) model 
on a kagom\'e lattice by Monte Carlo simulation 
to study the effects of a geometrical frustration, 
and the mechanism that generates an ordered state in a metallic system. 
The DESI model on the kagom\'e lattice is a frustrated metallic system 
due to an effective ferromagnetic interaction 
between localized spins caused by the double-exchange (DE) mechanism and 
a uniaxial anisotropy for the localized spins. 
A dodecagonal spin cluster (named dodecamer), 
which consists of twelve localized spins, appears at low temperature 
when the number of particles per site $n \simeq 1/3 \sim 1/2$.
Such a dodecamer order is driven by 
both the kinetic energy gain due to the DE mechanism 
and the geometrical frustration. 
We discuss that cluster orders, in general, may be a 
common feature in itinerant electron systems 
coupled with frustrated adiabatic fields.
}

\kword{%
dodecamer, cluster order, double-exchange model, kagom\'e lattice, 
frustration, metal, spin ice, Monte Carlo simulation
}

\begin{document}
\maketitle

%\section{Introduction}

Geometrically frustrated systems have been
an attractive subject due to their novel phenomena
after the predictions of disordered ground states 
in antiferromagnetic (AF) classical spin systems 
on geometrically frustrated lattices, 
i.e., a triangular\cite{Wannier_50}, 
a kagom\'e\cite{Syozi_51,Kano_53}, and
a pyrochlore lattice\cite{Anderson_56}.
These disordered systems have 
macroscopically degenerate ground states 
accompanied with a finite residual entropy.
Such a degeneracy can be lifted by additional effects,
which leads to a novel ground state.
For example, in frustrated quantum spin systems
\cite{Misguich_condmat}, 
quantum fluctuation sometimes lifts 
the degeneracy and induces a nonmagnetic ground state 
with a finite spin gap. 
One of the most well-known examples is 
a dimer singlet ground state 
observed in the Majumdar-Ghosh model\cite{Majumdar_69} and 
the Shastry-Sutherland model\cite{Shastry_81}, 
which probably reflects 
the short-range interaction 
between localized spins. 

The spin ice system 
in pyrochlore oxides\cite{Harris_97,Ramirez_99} 
provides us another insight on the lifting of the degeneracy. 
Localized spins in these systems 
have a strong uniaxial anisotropy 
in the $\langle 111 \rangle$ direction 
for each corner-shared tetrahedron 
that constitutes the pyrochlore lattice. 
The interaction between nearest-neighbor (n.n.) spins 
is ferromagnetic. 
Each tetrahedron has the ``two-in two-out'' spin structure 
at low temperature, 
i.e., two spins point inward for the tetrahedron and 
the other two outward 
due to the uniaxial anisotropy and the ferromagnetic interaction. 
This is often called an ice rule 
after the positions of hydrogen atoms in the ice. 
According to recent researches\cite{Byron_00,Bramwell_01,Melko_01}, 
a long-range dipolar interaction generates a cluster order 
satisfying the ice rule, 
which cannot be explained 
by the n.n. ferromagnetic Ising interaction. 
This indicates that the long-range interaction 
lifts the degeneracy and creates a cluster order. 

Considering these circumstances, 
it is highly significant to investigate various mechanisms 
for lifting the degeneracy due to the frustration. 
Naturally, frustrated electron systems are also 
expected to have a peculiar mechanism. 
As a candidate for electron systems, let us consider 
the double-exchange (DE) model \cite{Zener_51}.
In this model, conduction electrons interact with localized spins 
through the on-site Hund's-rule coupling $J_{H}$, 
which produces the effective ferromagnetic interaction 
between n.n. localized spins. 
However, the range of interaction 
between localized spins is determined by the kinetics 
of electrons \cite{Motome_01}. 
In the strong Hund's-rule coupling limit ($J_{H} \rightarrow \infty$),
spins of conduction electrons are parallel to localized spins 
and ferromagnetic domains are formed at low temperature.  
Since electrons are confined within the domains,
the electronic energy strongly depends on the sizes 
and the shapes of the domains.
Therefore, in order to describe the energy of the system 
as a function of spin configuration, 
it is necessary to take into account effective long-range
two- and/or multiple-spin interactions in each domain. 
Note that this is in contrast to the spin systems 
with the short-range interaction, 
where the energy depends on surface volumes of domain boundaries.
In this way, frustrated DE systems may have unique features
with respect to the mechanisms for lifting the degeneracy 
due to the kinetics of electrons. 

Following this idea, we have constructed 
the double-exchange spin ice (DESI) model on a kagom\'e lattice, 
as an example of frustrated DE models. 
The model has an effective ferromagnetic interaction 
due to the DE mechanism and  
a uniaxial anisotropy for localized spins 
as in the spin ice system. 
Therefore, this system has the frustration. 
In this study, we have investigated the low-temperature behavior 
of the DESI model using the Monte Carlo (MC) method. 
From MC calculations, we obtain the following: 
(1) A dodecagonal localized spin cluster, ``dodecamer'', 
is realized at low temperature, 
(2) the dodecamer phase exists in a wide doping region 
$n \simeq 1/3 \sim 1/2$, where $n$ is the number of particles per site, 
and (3) the dodecamer order is driven by 
both the kinetic energy gain due to the DE mechanism and the frustration.
On the analogy of the DESI system on the kagom\'e lattice, 
similar cluster orders might be generic features 
in frustrated electron systems.

%\section{Model}

Let us start from the DE model 
in the strong Hund's-rule coupling limit
\cite{Zener_51, Anderson-Hasegawa_55} 
\begin{equation}
H = -\sum_{\langle i,j \rangle} \left \{
\tilde{t}( \vec{S}_i , \vec{S}_j ) 
\tilde{c}_{i}^{\dagger} \tilde{c}_{j} + h.c.\right \}
- \mu \sum_{i} \tilde{c}_{i}^{\dagger} \tilde{c}_{i},
\label{eq:EffDEmodel}
\end{equation}
where $\tilde{c}_{i}^{\dagger}(\tilde{c}_{i})$ is 
an electron operator 
that creates (annihilates) an electron at site $i$ 
with a spin parallel to a localized spin $\vec{S}_i$. 
The effective hopping matrix element is defined as 
$\tilde{t}( \vec{S}_i , \vec{S}_j ) 
= t \cos \left( \theta_{ij}/2 \right) \exp(i \gamma_{ij})$, 
where $t$ represents the n.n. transfer integral, 
$\theta_{ij}$ is an angle between $\vec{S}_{i}$ and $\vec{S}_{j}$, 
and $\gamma_{ij}$ is a phase factor that creates the Berry phase 
at the noncoplanar configuration of spins. 
Hereafter, we set that $t$ is positive,
and localized spins are classical ones for simplicity. 
Note that the absolute value of 
$\tilde{t}( \vec{S}_i , \vec{S}_j )$ 
becomes a maximum when $\vec{S}_i$ and $\vec{S}_j$ are parallel, 
which leads to the effective ferromagnetic interaction 
between the n.n. spins to gain the kinetic energy locally. 
However, as described above, 
the electronic energy is determined by the kinetics of electrons, 
which depends on the sizes and shapes of ferromagnetic domains 
where electrons move easily. 
Thus, it is found that 
effective multiple interactions are important 
to determine the nature of the DE system. 

\begin{figure}
\begin{center}
\includegraphics[width=8cm,keepaspectratio]{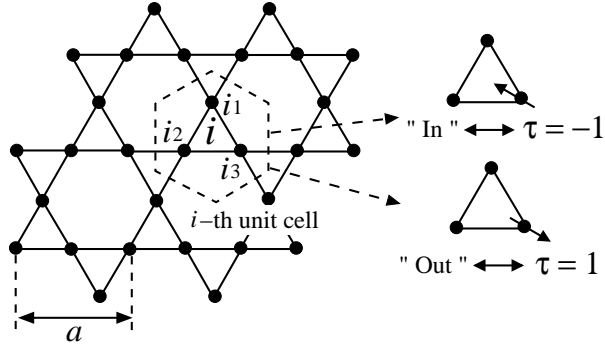}
\end{center}
\caption{
  DESI model on the kagom\'e lattice.
  The dashed line is an $i$-th unit cell
  composed of three sites ($i_1$,$i_2$ and $i_3$), 
  and $a$ is a lattice constant. 
  The direction of a localized spin is shown by arrows:
  inward (in-spin) or outward (out-spin) for an up-triangle 
  in the unit cell. $\tau$ is a pseudo-spin operator, where
  an out-(in-)spin corresponds to a pseudo-up (down) spin.} 
\label{fig:uniaxial}
\end{figure}

In this letter,
we consider the DE model on the kagom\'e lattice and 
assume a uniaxial anisotropy for the localized spin,
which is forced to point inward (in-spin) or 
outward (out-spin) for an up-triangle 
in a unit cell of the kagom\'e lattice 
[see Fig. \ref{fig:uniaxial}]. 
Spins are oriented 
within the kagom\'e plain, i.e., coplanar 
under this constraint. 
We consider the model without the phase 
factor $\gamma_{ij}$ in $\tilde{t}( \vec{S}_i , \vec{S}_j )$. 
An in-spin is energetically preferable 
next to an out-spin on each triangle 
due to the DE mechanism. 
This situation is similar to the spin ice system.\cite{Ramirez_99,Harris_97} 
Therefore, we call the model the DESI model. 
We define a pseudospin at site $i$,
\begin{equation}
\tau_{i} = 
\begin{cases}
+1 & \text{($\vec{S}_{i}$ is an out-spin)}, \\
-1 & \text{($\vec{S}_{i}$ is an in-spin)}.
\end{cases}
\end{equation}
In this representation, the effective ferromagnetic interaction 
between n.n. localized spins is regarded as 
the AF interaction between n.n. pseudospins. 
Thus, it is naturally considered that each triangle has 
a ``two-down one-up'' or ``one-down two-up'' pseudospin structure 
at low temperature, which corresponds to 
a ``two-in one-out'' or ``one-in two-out'' localized spin structure, 
respectively. 
Hereafter, we use the pseudospin picture \{$\tau_{i}$\}
instead of the bare localized spin picture \{$\vec{S}_{i}$\}. 
One may think that the DESI model on the kagom\'e lattice 
can be mapped onto the AF Ising model on the same lattice, 
which has a disordered ground state 
with a macroscopic degeneracy.\cite{Syozi_51,Kano_53} 
However, the nature of the former system should be determined 
by the kinetics of electrons, i.e., 
effective long-range interaction, and 
it is expected that the behavior of the DESI model may be 
quite different from the frustrated Ising spin system
with the short-range interaction. 

We have performed MC calculations\cite{Yunoki_98} 
to study the thermodynamics of 
the system described by the Hamiltonian (\ref{eq:EffDEmodel}) 
at finite temperatures. 
We typically run 100,000 MC steps 
for measurement after 10,000 thermalization steps. 
We applied the Metropolis algorithm for the updates of 
spin configurations.\cite{Metropolis_53}

%\section{Results}

\begin{figure}
\begin{center}
\includegraphics[width=8cm,keepaspectratio]{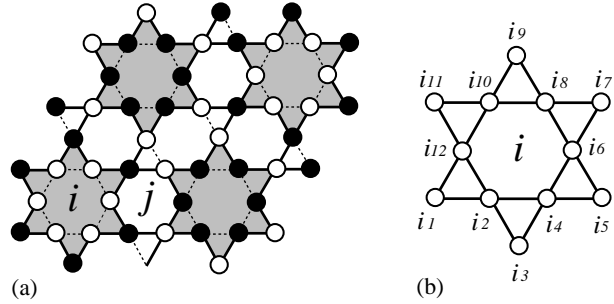}
\end{center}
\caption{
(a) Snapshot of localized spins at $k_B T/t=0.01$ and $\mu / t=0$. 
In-(out-)spins are represented by open (closed) circles.
An in-spin and an out-spin are connected by a solid line, 
while the nearest-neighbor pairs of in-(out-)spins 
are connected by a dashed line. 
Electrons can move easily through solid lines. 
Each hexagon is labeled $i$, $j$ and others. 
The shadowed dodecagonal spin cluster is a dodecamer.
(b) A dodecamer consists of twelve localized spins. 
Labels $i_1,\dots,i_{12}$ are site numbers on the $i$-th dodecamer.}
\label{fig:snapshot}
\end{figure}

In order to investigate the features of the localized spins, 
we have calculated a pseudospin structure factor defined as 
\begin{equation}
T_{q}^{\alpha \beta}=\frac{1}{N_{\rm c}}\sum_{i,j}e^{{\rm i} \vec{q} \cdot 
(\vec{r}_{i_{\alpha}}-\vec{r}_{j_{\beta}})} 
\langle \tau_{i_{\alpha}} \tau_{j_{\beta}} \rangle,
\label{eq:tau_corr}
\end{equation}
where $N_{\rm c}$ is the number of unit cells on the kagom\'e lattice, 
$i_{\alpha}$ and $i_{\beta}$ ($\alpha$ and $\beta$ = 1, 2, or 3) 
indicate three sites in an $i$-th unit cell 
shown in Fig. \ref{fig:uniaxial}, 
and the summation for $i$ and $j$ extends over all unit cells. 
$ \langle \cdots \rangle $ represents the thermal average. 
Given the chemical potential $\mu / t=0$ 
and $N_{\rm c}= 4 \times 4$ (= 48 sites) 
and $8 \times 8$ (= 192 sites), 
results at sufficiently low temperature ($k_{B}T/t = 0.01$) 
are as follows: 
(1) Maximum eigenvalues of $T_{q}^{\alpha \beta}$ are observed at 
$\vec{q}_1={}^{t}(3 \pi / 2a , \pi / 2a)$ and 
$\vec{q}_2={}^{t}(\pi / 2a , 3 \pi / 2a)$.
(2) Corresponding eigenmodes represent 
a two-in one-out or one-in two-out spin structure. 
These results mean that 
the translational symmetry of the lattice breaks
at low temperature. 
Such a symmetry breaking is also observed simultaneously 
in a snapshot of localized spins [see Fig. \ref{fig:snapshot}(a)], and
the periodicity of spins is almost consistent with 
that represented by $\vec{q}_1$ and $\vec{q}_2$.
Note that there are two kinds of bond: 
a bond between in- and out-spins 
[solid line in Fig. \ref{fig:snapshot}(a)] 
and that between same spins 
[dashed line in Fig. \ref{fig:snapshot}(a)]. 
Electrons can easily move through the former bond 
due to a larger $\tilde{t}( \vec{S}_i , \vec{S}_j )$ value. 
Connecting bonds of the former type, the kagom\'e lattice 
can be tiled by the dodecagonal spin cluster 
which we call a ``dodecamer'' after twelve localized spins, 
as shown in Fig. \ref{fig:snapshot}(a).

The existence of the dodecamers can be read off 
using an order parameter defined by 
\begin{equation}
d_{i} \equiv \frac{1}{18} 
\left( 
- \sum_{n=1}^{12} \tau_{i_{n}} \tau_{i_{n+1}}
+ \sum_{m=1}^{6} \tau_{i_{2m}} \tau_{i_{2m+2}}
\right),
\label{eq:order_param}
\end{equation}
where a site $i_{n} (n=1,2,\cdots, 12)$ is indicated 
in Fig. \ref{fig:snapshot} (b) and 
we set $\tau_{i_{13}} = \tau_{i_{1}}$ and $\tau_{i_{14}} = \tau_{i_{2}}$ 
from the periodicity of the lattice. 
Note that the order parameter $d_{i}$ is a unity 
when the dodecamer exists. 
The structure factor of a correlation function  
$\langle d_{i}d_{j} \rangle$
is defined by
\begin{equation}
D_{q} = \frac{1}{N_{\rm c}} \sum_{i,j} 
        e^{{\rm i} \vec{q} \cdot (\vec{r}_{i}- \vec{r}_{j})}
        \langle d_{i} d_{j} \rangle ,
\label{eq:Dq}
\end{equation}
where $\sum$ represents the summation 
for any pairs of the order parameters 
shown in Fig. \ref{fig:snapshot} (a).

\begin{figure}
\begin{center}
\includegraphics[width=8cm,keepaspectratio]{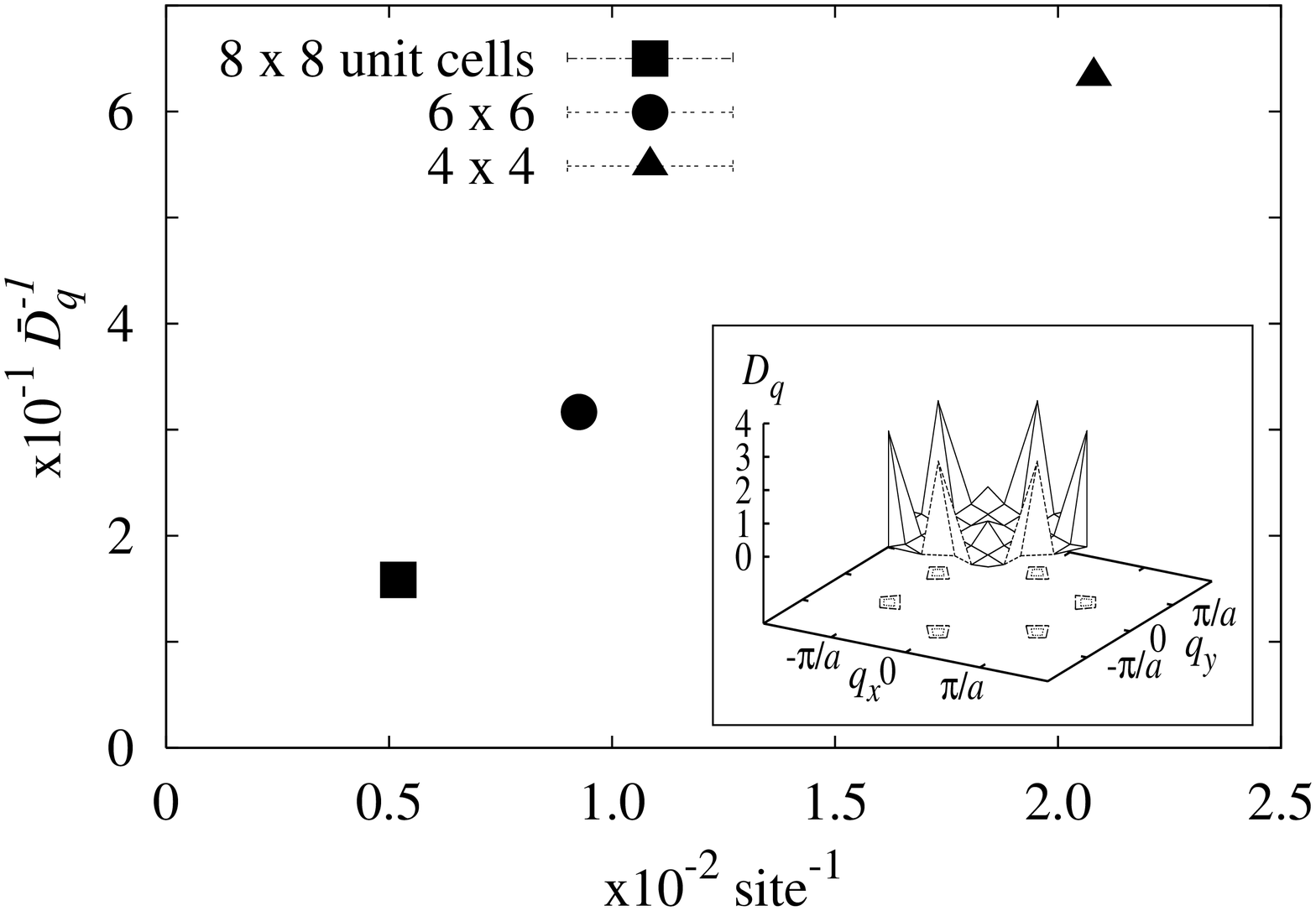}
\end{center}
\caption{System size dependence of $\bar{D}_q$ 
at $k_B T / t=0.006$ and $\mu / t = 0$, 
where $\bar{D}_q \equiv 
\{ D_{Q_1}+D_{Q_2}+D_{Q_3} \}/3$. 
$\blacksquare$, $\bullet$, and $\blacktriangle$ 
represent 8 $\times$ 8 (= 192 sites), 
6 $\times$ 6 (= 108 sites) and 4 $\times$ 4 (= 48 sites)
unit cells, respectively.
Error bars are within the sizes of the symbols. 
It is found that $\bar{D}_q$ grows proportional to $N_{\rm c}$.
The inset is the dodecamer structure factor 
with the lattice size $N_{\rm c} = 6 \times 6$.} 
\label{fig:Dq_size_dpnd}
\end{figure}

The results with the lattice size 
$N_{\rm c} = 6 \times 6$ at $k_B T / t =0.006$ and $ \mu / t =0$ 
are shown in the inset of Fig. \ref{fig:Dq_size_dpnd}. 
There are three independent 
peaks at $\vec{Q}_{1}={}^t(0,2 \pi / \sqrt{3} a)$, 
$\vec{Q}_{2}={}^t( \pi / a, \pi / \sqrt{3} a)$ 
and $\vec{Q}_{3}={}^t( \pi / a, - \pi / \sqrt{3} a)$, 
which is consistent with the periodicity of the dodecamer order state 
shown in Fig. \ref{fig:snapshot} (a). 

The system size dependence of the averaged structure factor 
$\bar{D}_q \equiv \{ D_{Q_{1}}+D_{Q_{2}}+D_{Q_{3}} \}/3$ 
for $N_{\rm c} = 4 \times 4$, $6 \times 6$ and $8 \times 8$ 
is shown in Fig. \ref{fig:Dq_size_dpnd}. 
The result indicates that the structure factor 
is proportional to the system size $N_{\rm c}$. 
Thus, we concluded that the dodecamer order survives 
even in the thermodynamic limit.

\begin{figure}
\begin{center}
\includegraphics[width=8cm,keepaspectratio]{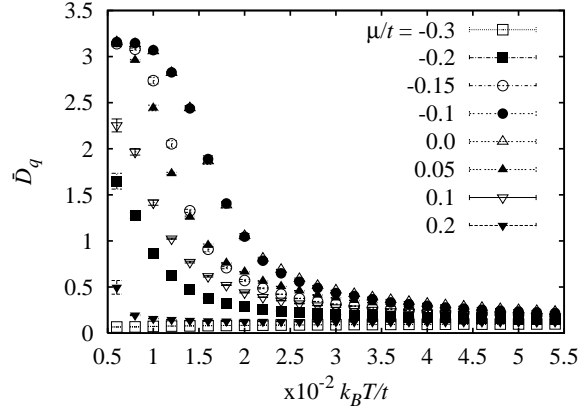}
\end{center}
\caption{
Temperature dependence of $\bar{D}_q$ at $\mu / t $ = -0.3 ($ \square $), 
 -0.2 ($ \blacksquare $), -0.15 ($ \circ $), -0.1 ($ \bullet $),
0.0 ($ \triangle $), 0.05 ($ \blacktriangle $), 0.1 ($ \triangledown $) 
and 0.2 ($ \blacktriangledown $) 
when the system size $N_{\rm c}$ is 6 $\times 6$ (= 108 sites).}
\label{fig:Dq_vs_T_at_some_mu}
\end{figure}

\begin{figure}
\begin{center}
\includegraphics[width=8cm,keepaspectratio]{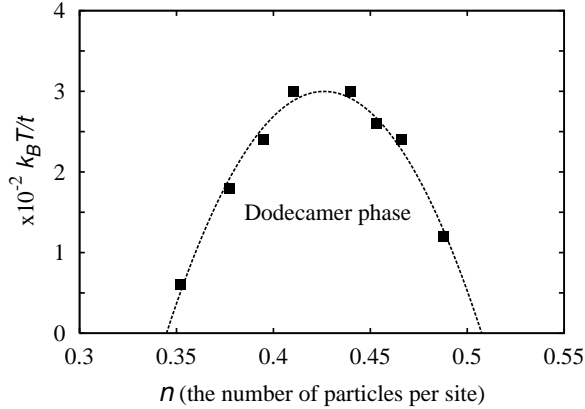}
\end{center}
\caption{
Estimated phase diagram with $N_{\rm c} = 6 \times 6$ 
($= 108$ sites) as a function of the number of 
particles per site $n$ and the temperature. 
The dodecamer phase can exist in the 
region $n \ \simeq \ 1/3 \ \sim \ 1/2 $. }
\label{fig:phase_diagram}
\end{figure}

To see the stability of the dodecamer order, 
the temperature dependence of $\bar{D}_q$ 
with various chemical potentials 
($\mu / t$ = -0.3, -0.2, -0.15, -0.1, 0.0, 0.05, 0.1 and 0.2) 
at $k_B T/t \leq 0.054$ 
has been calculated.
As shown in Fig. \ref{fig:Dq_vs_T_at_some_mu}, 
several $\bar{D}_q$'s grow rapidly at low temperature,
which roughly indicates a phase transition 
from the disordered state to the dodecamer state.
Note that the constraints for localized spins, 
i.e., two-in one-out or one-in two-out, are satisfied 
in this temperature range.
We define the pseudotransition temperature 
from the deviation of the data 
in Fig. \ref{fig:Dq_vs_T_at_some_mu} from the Curie-Weiss law. 
Such a transition temperature is not 
rigorous but at least gives an upper limit. 
An estimated phase diagram is summarized 
in Fig. \ref{fig:phase_diagram}.
We expect that the dodecamer order exists
in the doping region 
$n \simeq 1/3 \sim 1/2 \ (-0.3 \lesssim \mu \lesssim 0.2)$. 

\begin{figure}
\begin{center}
\includegraphics[width=8cm,keepaspectratio]{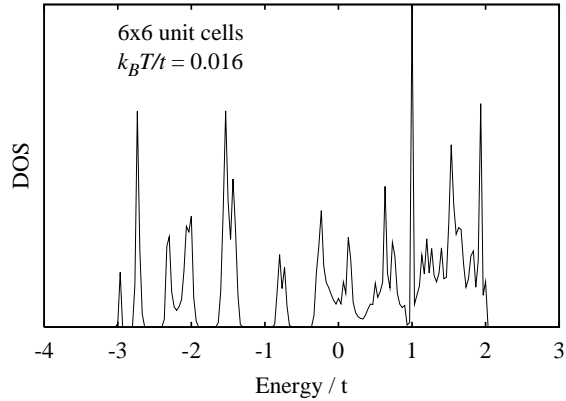}
\end{center}
\caption{Density of states (DOS) at 
$N_{\rm c}= 6 \times 6 \ (108 \ {\rm sites})$ and $k_B T/t=0.016$. 
The dodecamer order appears at $-0.3 \lesssim \mu \lesssim 0.2$.}
\label{fig:DOS}
\end{figure}

Let us consider the origin of the dodecamer order state.
We consider a few scenarios:
(1) the formation of an electron hopping path 
where the kinetic energy of electrons can be gained 
in terms of the DE mechanism and 
(2) the stabilization caused by an energy gap around the Fermi level 
breaking the translational symmetry of the lattice in some manners. 
The former case corresponds to a metallic state, 
and the latter case, an insulating state. 
To clarify this point, 
we have investigated the density of states (DOS).
The DOS at $N_{\rm c} = 6 \times 6$ and $k_B T / t =0.016$ 
is shown in Fig. \ref{fig:DOS}. 
The result indicates that the energy gap is not opened 
around the chemical potentials ($-0.3 \lesssim \mu/t \lesssim 0.2$)
when the dodecamer state is realized, 
which means that the dodecamer state corresponds to a metallic state 
in terms of the conventional band picture.
This result excludes the latter possibility and is reasonable, 
because the DESI model originally has the only DE mechanism 
that prevents insulating states. 
This is also consistent with the result that 
the stability of the dodecamer order 
is not sensitive to the shape of the Fermi surface 
indicated by the existence of the wide doping region 
for the dodecamer state. 
Note that the two-in one-out or one-in two-out structure 
has been maintained in the dodecamer state.  
This finding indicates that the ferromagnetic interaction, that is,
the DE mechanism due to the metallic nature of electrons 
is important even in the dodecamer phase. 
This strongly indicates that 
the dodecamer cluster is driven by 
both the frustration and the kinetic energy gain 
due to the DE mechanism. 

%\section{Consideration}

In conclusion, 
the dodecamer order appears at low temperature in the DESI model 
on the kagom\'e lattice 
because of the kinetic energy gain due to the DE mechanism 
and the frustration of the model. 
The DESI model can be realized in other lattices, for example, 
the pyrochlore lattice. 
Thus, it is natural to expect 
the cluster order state in the DESI model 
not only on the kagom\'e lattice 
but also on the pyrochlore lattice. 
Cluster orders may exist in real materials 
which have the pyrochlore structure, the DE mechanism, and the frustration. 
One of the possibilities is that in a pyrochlore oxide 
$R_{2}{\rm Mo_{2}O_{7}}$,
where $R$ is a trivalent rare-earth ion (Nd, Sm, Gd, Tb, Dy, Yb) and Y.
\cite{Greedan_87, Ali_89, Katsufuji_00, Moritomo_01, Taguchi_01} 
There is a crossover from the 
ferromagnetic metallic (FM) phase to the spin-glass (SG)
phase as the mean ionic radius $R$ decreases, i.e., 
the bandwidth control\cite{Katsufuji_00,Moritomo_01}.  
Since the magnetoresistance effect supports that 
the FM state is caused by the DE mechanism 
between $d$ electrons on ${\rm Mo^{4+}}$,
this series might be a realization of the DESI model
due to both the geometry of the pyrochlore lattice 
and the strong anisotropy of spins.  
In particular, the FM state near the phase boundary 
has a character of the SG state indicated by ordinary 
and anomalous Hall coefficients \cite{Katsufuji_00}
and an elastic neutron scattering measurement also 
indicates a short-range spin order 
in ${\rm Y_2 Mo_2 O_7}$\cite{Gardner_99}, 
which may be explained by considering cluster orders.
If the uniaxial anisotropy of spins in our model is weakened, 
i.e., spins change from Ising to Heisenberg, 
it is obvious that the system has the FM ground state, 
which corresponds to ${\rm Nd_2 Mo_2 O_7}$. 
In this way, we may explain the transition changing the strength 
of the uniaxial anisotropy and anomalous behaviors 
observed in this series.

In terms of cluster formation in frustrated systems, 
the effective interaction due to the kinetics of electrons 
in the DESI system corresponds to the long-range dipolar interaction 
in the spin ice system. 
Thus, we expect that the long-range nature is important to create 
cluster orders in both frustrated spin and electron systems. 
Although two-body (short-range) interactions, 
such as the n.n. Ising interaction, 
are usually sufficient to create ordered states in nonfrustrated systems, 
these interactions are not sufficient to lift the degeneracy 
in the frustrated systems. 
Thus, other effects, which can be neglected 
in conventional nonfrustrated systems, 
are important in frustrated systems. 
In the DESI model, 
the kinetic energy gain corresponds to such effects.
In other frustrated electron systems,
the formation of similar clusters 
where electrons can move easily 
is also expected to gain the kinetic energy. 
For example, an electron-lattice coupled system is the case, 
where we may find the lattice-distorted cluster. 
Even in frustrated systems 
with the short-range electron-electron interaction 
such as the Hubbard model, 
similar cluster orders can exist in the same manner. 
Since studies in the Hubbard model on the
frustrated lattice are limited\cite{Imai_03},   
further developments, particularly 
apart from the half-filling case, are desired.

We would like to thank H. Tsunetsugu 
and N. Kawakami for helpful suggestions. 
The authors also thank M. Tsutsui and S. Katsurada 
for computational support. 
The numerical computations have been performed 
mainly using the facilities of the AOYAMA+ project. 
This work was partially supported by a Grant-in-Aid for 21st COE program 
from the Ministry of Education, Culture, Sports, Science 
and Technology of Japan.

\end{document}